# Mechanisms of Electromechanical Coupling in Strain Based Scanning Probe Microscopy

Qian Nataly Chen [1,*], Yun Ou [2,*], Feiyue Ma [1], and Jiangyu Li [1,†]

1   Department of Mechanical Engineering, University of Washington, Seattle, WA 98195-2600, USA

2   Faculty of Materials, Optoelectronics and Physics, and Key Laboratory of Low Dimensional Materials & Application Technology of Ministry of Education, Xiangtan University, Xiangtan 411105, Hunan, China

**Abstract**

Electromechanical coupling is ubiquitous in nature and underpins the functionality of materials and systems as diverse as ferroelectric and multiferroic materials, electrochemical devices, and biological systems, and strain-based scanning probe microscopy (s-SPM) techniques have emerged as a powerful tool in characterizing and manipulating electromechanical coupling at the nanoscale. Uncovering underlying mechanisms of electromechanical coupling in these diverse materials and systems, however, is a difficult outstanding problem, and questions and confusions arise from recent experiment observations of electromechanical coupling and its apparent polarity switching in some unexpected materials. We propose a series of s-SPM experiments to identify different microscopic mechanisms underpinning electromechanical coupling, and demonstrate their feasibility using three representative materials. By employing a combination of spectroscopic studies and different modes of s-SPM, we show that it is possible to distinguish electromechanical coupling arising from spontaneous polarization, induced dipole moment, and ionic Vegard strain, and this offer a clear guidance on using s-SPM to study a wide variety of functional materials and systems.

[*] Those authors contribute equally to the work.
[†] Author to whom the correspondence should be addressed to; email: jjli@uw.edu

Since the pioneering development of piezoresponse force microscopy (PFM) in the 1990s [1-3], strain-based scanning probe microscopy (s-SPM) techniques have emerged as a powerful tool in characterizing and manipulating functional materials and structures at the nanoscale. For ferroelectrics, PFM has been widely used to image local domain characteristics, providing direct experimental observations on switching and fatigue, domain-defect interactions, and nucleation [4-6]. For lithium ion batteries and solid oxide fuel cells, electrochemical strain microscopy (ESM) has enabled probing of electrochemistry at the nanoscale, offering much higher spatial resolution compared to conventional current based techniques [7-9]. Recently, piezomagnetic force microscopy (PmFM) has also been developed to image magnetic materials and structures, making it possible to probe local magnetic ordering quantitatively using SPM [10, 11]. While applicable to a wide range of material systems and phenomena, s-SPM techniques are particularly suitable for probing electromechanical coupling at the nanoscale, which is ubiquitous in nature and underpins the functionality of materials and systems as diverse as ferroelectrics and multiferroic materials, electrochemical devices, and biological systems [4]. Nevertheless, the underlying mechanisms in these wide varieties of materials and systems could be vastly different, and distinguishing electromechanical mechanisms in these diverse materials and systems is a difficult outstanding problem. We seek to address this issue in this letter.

To this end, we highlight recent experimental observations of electromechanical coupling and its apparent polarity switching in two unexpected materials, one in glass [12] and the other in silicon [13]. Neither material is piezoelectric or ferroelectric, yet they not only respond to local electrical excitation applied via SPM probe, but the phases of their strain responses also appear to be switchable by an electric field, resulting in hysteresis and butterfly loops that resemble characteristics of classical ferroelectrics. These experiments cast doubts on PFM hysteresis and butterfly loops as signatures of ferroelectric switching, and raise questions on recent reports of biological ferroelectricity [14, 15]. In addition, since electromechanical coupling is increasingly applied to probe electrochemical systems via compositional Vegard strain [16, 17], it becomes important to be able to distinguish piezoelectric and electrochemical strains in s-SPM imaging as well. In order to address these issues, we examine three different classes of materials using s-SPM techniques in this letter, and we propose a series of s-SPM experiments to pin down the underlying mechanisms of their electromechanical coupling.

We choose three representative material systems to study, including lead zirconate titanate (PZT) thin film, a classical ferroelectric, lithium iron phosphate (LFP) film, an electrochemical material with large Vegard strain [8], and soda-lime glass, which is an amorphous ionic system [12]. The electromechanical imaging of these materials via s-SPM all operates under the same principle – an AC voltage is applied to the specimen through the SPM probe in contact mode, exciting a vibration of the cantilever due to the electromechanical coupling of the specimen. The vertical deflection and lateral twisting of the cantilever due to the out-of-plane normal and in-plane shear strains of the specimen are then measured by photodiode, yielding both phase and amplitude data of the vertical and lateral s-SPM responses that reflect the underlying electromechanical coupling of the specimen. Typical mappings of vertical s-SPM amplitude and phase corrected using the damped harmonic oscillator model [18] are shown in Fig. 1, exhibiting no clear distinctions of electromechanical responses among these three very different systems, except that the mappings of PZT show polycrystalline structure, while those of glass reveal no topographic features due to its amorphous nature. The question then is can we distinguish the microscopic mechanisms of electromechanical coupling using s-SPM, especially when a material is new or unknown?

Classical ferroelectrics exhibit piezoelectric strain biased by their spontaneous polarization [19], as schematically shown in Fig. 2a. It thus exhibits bipolar characteristics that can be switched by an external DC voltage. Electrochemical systems, on the other hand, may possess Vegard strain that depends only on the ionic composition, and thus is nonpolar in nature [8], as shown in Fig. 2b. Due to the nonpolar characteristic of electrochemical Vegard strain, DC field can only manipulate the ionic concentration and correspondingly the magnitude of its strain response, not its phase. This provides us a method to distinguish dipolar piezoelectric strain and nonpolar electrochemical strain via s-SPM spectroscopy technique – a sequence of DC voltages is applied to the specimen, with an AC voltage superimposed to excite electromechanical vibration of the sample. Both phase and amplitude of the vibration are recorded as a function of DC voltage, and if the phase is switched upon reversed DC, then the strain could be dipolar piezoelectric; otherwise it should be nonpolar electrochemical. This is indeed what we observe in PZT and LFP, respectively, as shown in Fig. 3. Both hysteresis and butterfly loops characteristic of classical ferroelectric switching are observed in PZT, which have been reported in a wide range of ferroelectric materials and structures. The sharp rise near coercive voltage is an indication of domain switching, resulting in a large extrinsic contribution to the strain. The phase of LFP, on the other hand, only shows small variations with respect to DC without switching, and its amplitude is anti-symmetric instead of symmetric with respect to the DC, consistent with compositional dependence of nonpolar Vegard strain. Similar characteristics have also been observed in other electrochemical systems dominated by Vegard strain. Thus the lack of phase reversal under s-SPM spectroscopic study is a defining characteristic of electrochemical Vegard strain, and this can be used to distinguish dipolar piezoelectric and nonpolar electrochemical strain.

What is surprising though is that soda-lime glass also exhibits hysteresis and butterfly loops similar to those of classical ferroelectrics, as shown in Fig. 3c. This is unexpected, as it is neither piezoelectric nor ferroelectric. The reversibility of the phase suggest that the strain response in glass is dipolar in nature, and thus it has been proposed that the electromechanical coupling could be related to dipoles induced by ionic motion under external electric field [12], as schematically shown in Fig. 2c. The question then is how we distinguish strain responses arising from induced versus spontaneous dipoles, both of which exhibit reversibility of phase and similar hysteresis loops, as seen in Fig. 3. To this end, we recognize that electromechanical strains from both induced and spontaneous polarizations are electrostrictive in nature, and thus are quadratic to the polarization, $\varepsilon \propto (P_s + \chi E)^2 = P_s^2 + 2\chi P_s E + \chi^2 E^2$, where $P_s$ is spontaneous polarization, and $\chi E$ is induced one. For classical ferroelectrics with large spontaneous polarization under small AC fields, the strain is predominantly linear to the AC field. On the other hand, for materials with small spontaneous polarization in comparison with the induced one, significant strain responses quadratic to the AC field would be expected. Thus, by comparing first and second harmonic responses under s-SPM, we could distinguish induced dipolar response from classical ferroelectrics. In order to acquire both first and second harmonic responses of the specimen, the resonant frequency of the cantilever-specimen system, $\omega_0$, is identified first, and then AC excitation is applied around $\omega_0$ (first harmonic) or $\omega_0/2$ (second harmonic), while the response is measured around $\omega_0$ through lock-in amplifier [20]. The comparisons between first and second harmonic responses under vertical s-SPM of these three classes of materials are shown in Fig. 4, and the distinction is evident. For PZT, the response is predominantly linear, as exhibited by one order of magnitude higher first harmonic response (0.19±0.045) compared to second harmonic one (0.018±0.0053), averaged over 9 different spatial points for each sample, with the amplification by quality factor corrected using the damped harmonic oscillator model [18]. For glass, on the other hand, the first (0.041±0.0023) and second (0.05±0.00067) harmonic responses sampled over 9 points are comparable. In fact the second harmonic response is slightly higher and has much smaller standard deviation, suggesting substantial quadratic contribution to strain, which is expected from induced dipole moments. Interestingly, LFP also shows significant quadratic strain, though here the second harmonic response (0.04±0.0068) is slightly smaller than the first harmonic one (0.05±0.016). As such, by examining first and second harmonic responses under s-SPM, it is possible to distinguish electrostrictive strain arising from spontaneous and induced polarizations, and in combination with spectroscopic switching, electrochemical strain can also be identified.

The electromechanical coupling in glass deserves further discussion, as it has caused much confusion recently. It exhibits dipolar electrostrictive strain response, though the dipoles are induced by ionic motion under an electric field. As such, it possesses mixed characteristics of both polar and electrochemical systems. First of all, the coercive field is not well defined in glass, as higher electric fields would induce larger extents of ionic diffusion, and thus opens the hysteresis loop further with increased maximum voltage.

This is indeed observed in glass, as shown in Fig. 5(a), and a similar observation has been made by Proksch [12]. For PZT, the hysteresis loop would not open further once the coercive field is reached (Fig. 5b). Another consequence is the time dependence of strain response in glass, which is limited by ionic transport and is much slower than the dynamics of intrinsic polarization. This is evident in butterfly loops of glass and PZT under different periods of DC cycles, as shown in Fig. 5(cd). Glass exhibits strong time dependence, with longer periods resulting in higher strain responses, consistent with longer range of ionic redistribution and thus larger induced dipoles, while PZT is insensitive to the time period. Another implication is that the induced dipoles in an amorphous system such as glass tend to have rather high macroscopic symmetry, aligned with the applied electric field, and thus the lateral shear response in plane, if any, will be much smaller than vertical one out of plane. This is in contrast to typical ferroelectrics such as PZT, for which substantial lateral responses comparable to vertical ones are expected due to lower symmetry. This is also confirmed by our data, as shown in Fig. 5(ef). The lateral response of soda-lime glass is one order of magnitude smaller than the vertical one, in both first (0.0058±0.0014) and second (0.0033±0.00075) harmonic modes, while PZT shows first harmonic lateral responses (0.12±0.01) comparable to that of the vertical one. Thus, by examining time-dependence of spectroscopic data, and by comparing lateral and vertical responses in first and second harmonics, electromechanical coupling arising from dipoles induced by ionic motion can be identified unambiguously by s-SPM.

We close by reaffirming s-SPM techniques as a powerful tool in probing electromechanical coupling of functional materials and systems, though caution must be exercised on different microscopic mechanisms underlying the electromechanical coupling. By employing a combination of spectroscopic studies and their time dependences, examining first and second harmonic strain responses, and comparing vertical and lateral amplitudes, it is possible to distinguish electromechanical coupling arising from spontaneous polarization, induced dipole moment, and ionic Vegard strain, and these offer a clear guidance on using s-SPM techniques to study a wide range of functional materials and systems.

**Acknowledgement**

We acknowledge partial support of NSF (CMMI-1100339) and NSFC (Approval Nos. 11102175 and 11372268).

# Reference

[1]     K. Franke, J. Besold, W. Haessler, and C. Seegebarth, Surf. Sci. **302**, L283 (1994).
[2]     O. Kolosov, A. Gruverman, J. Hatano, K. Takahashi, and H. Tokumoto, Phys. Rev. Lett. **74**, 4309 (1995).
[3]     A. Gruverman, O. Auciello, and H. Tokumoto, J. Vac. Sci. Technol. B **14**, 602 (1996).
[4]     S. V. Kalinin, B. J. Rodriguez, S. Jesse, E. Karapetian, B. Mirman, E. A. Eliseev, and A. N. Morozovska, Annu. Rev. Mater. Res. **37**, 189 (2007).
[5]     H. F. Yu, H. R. Zeng, R. Q. Chu, G. R. Li, and Q. R. Yin, J. Inorg. Mater. **20**, 257 (2005).
[6]     N. Balke, I. Bdikin, S. V. Kalinin, and A. L. Kholkin, J. Am. Ceram. Soc. **92**, 1629 (2009).
[7]     N. Balke *et al.*, Nat. Nanotechnol. **5**, 749 (2010).
[8]     Q. N. Chen, Y. Y. Liu, Y. M. Liu, S. H. Xie, G. Z. Cao, and J. Y. Li, Appl. Phys. Lett. **101** (2012).
[9]     J. Zhu, L. Lu, and K. Y. Zeng, ACS Nano **7**, 1666 (2013).
[10]    Q. N. Chen, F. Y. Ma, S. H. Xie, Y. M. Liu, R. Proksch, and J. Y. Li, Nanoscale **5**, 5747 (2013).
[11]    A. Eshghinejad, W. I. Liang, Q. N. Chen, F. Ma, Y. Liu, S. Xie, Y. H. Chu, and J. Li, J. Appl. Phys.  (to be published).
[12]    R. Proksch, J. Appl. Phys.  (to be published).
[13]    J. S. Sekhon, L. Aggarwal, and G. Sheet, arXIV:1401.2512.
[14]    Y. M. Liu, Y. J. Wang, M. J. Chow, Q. N. Chen, F. Y. Ma, Y. H. Zhang, and J. Y. Li, Phys. Rev. Lett. **110**, 168101 (2013).
[15]    Y. M. Liu, Y. H. Zhang, M. J. Chow, Q. N. Chen, and J. Y. Li, Phys. Rev. Lett. **108** (2012).
[16]    S. Kalinin, N. Balke, S. Jesse, A. Tselev, A. Kumar, T. M. Arruda, S. L. Guo, and R. Proksch, Mater. Today **14**, 548 (2011).
[17]    A. N. Morozovska, E. A. Eliseev, A. K. Tagantsev, S. L. Bravina, L. Q. Chen, and S. V. Kalinin, Phys. Rev. B **83** (2011).
[18]    S. H. Xie, A. Gannepalli, Q. N. Chen, Y. M. Liu, Y. C. Zhou, R. Proksch, and J. Y. Li, Nanoscale **4**, 408 (2012).
[19]    J. Y. Li, Y. M. Liu, Y. H. Zhang, H. L. Cai, and R. G. Xiong, Phys Chem Chem Phys **15**, 20786 (2013).
[20]    Y. Kim *et al.*, ACS Nano **5**, 9104 (2011).

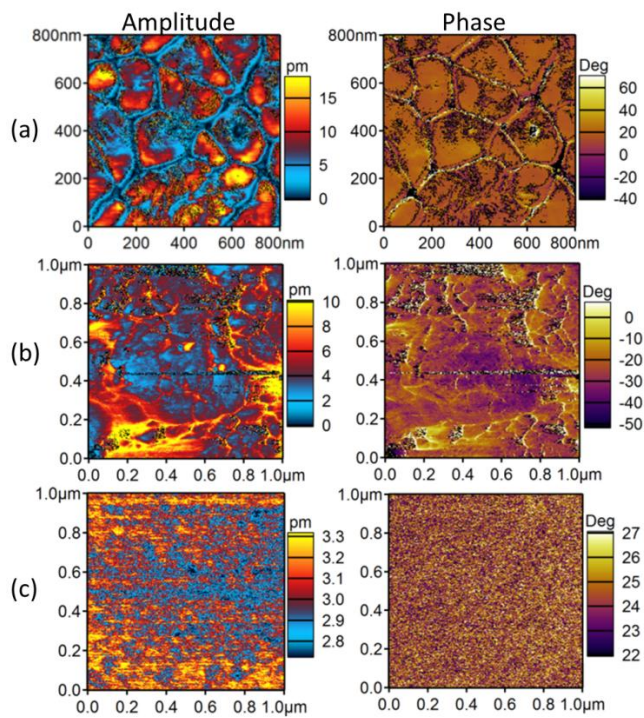

**Fig. 1** Vertical s-SPM amplitude and phase mappings of electromechanical responses of three representative material systems; (a) PZT; (b) LiFePO$_4$; and (c) soda-lime glass.

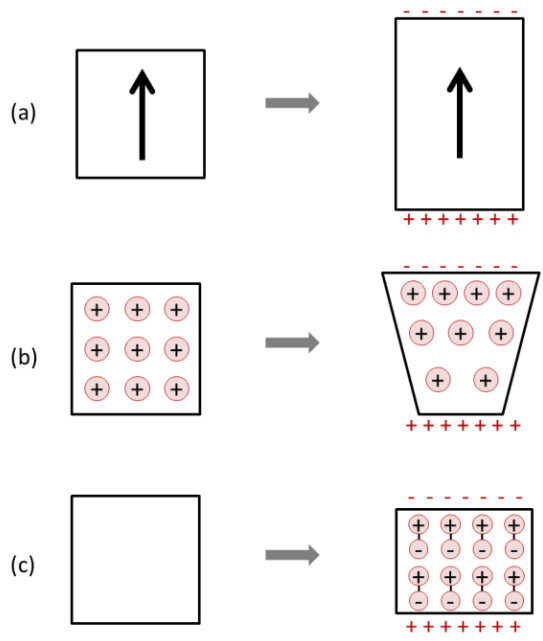

**Fig. 2** Schematics of three microscopic mechanisms of electromechanical coupling; (a) piezoelectric; (b) electrochemical; and (c) electrostrictive.

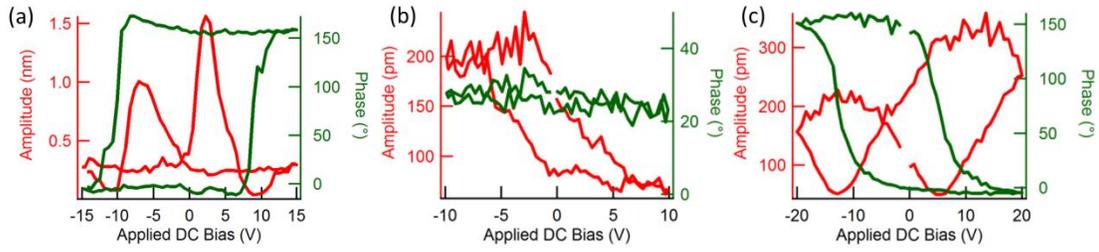

**Fig. 3** Phase-voltage and amplitude-voltage loops of three representative material systems; (a) PZT; (b) LFP; and (c) soda-lime glass.

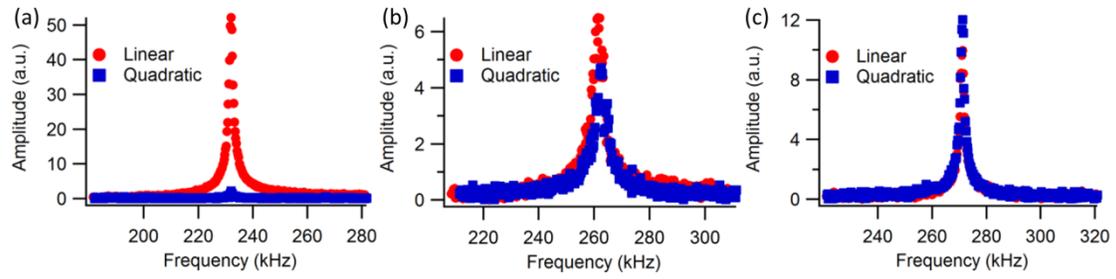

**Fig. 4** First and second harmonic responses of three representative material systems in vertical s-SPM; (a) PZT; (b) LFP; and (c) soda-lime glass.

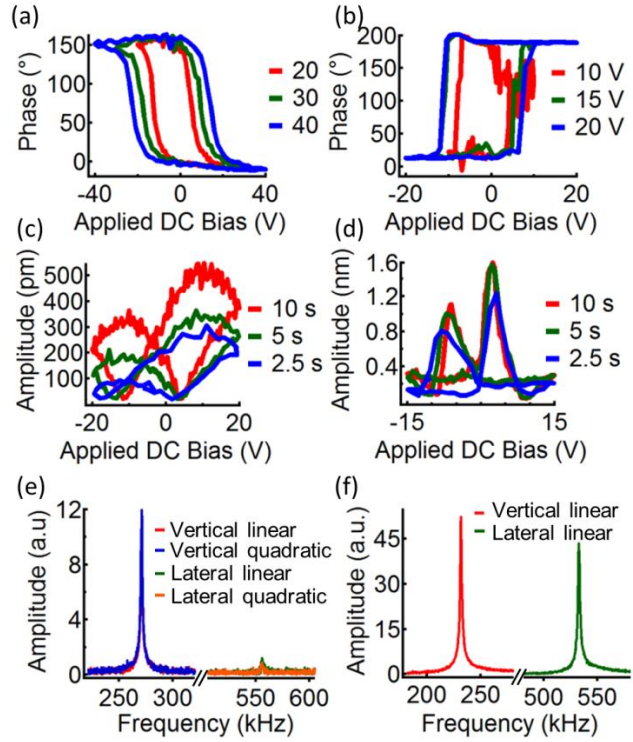

**Fig. 5** Additional comparison of spontaneous and induced ionic polarization; hysteresis loops of (a) soda-lime glass and (b) PZT in vertical s-SPM under different maximum DC voltage; butterfly loops of (c) soda-lime glass and (d) PZT in vertical s-SPM under different cycling period; and vertical and lateral s-SPM response of (e) soda-lime glass and (f) PZT.

11